\pdfoutput=1

\documentclass[twocolumn,final]{svjour3}

\smartqed

\usepackage{graphicx,amssymb}
\usepackage{epsf,psfig}

\journalname{Nonlinear Dynamics}

\begin{document}

\title{Investigating the nature of motion in 3D perturbed elliptic \\
 oscillators displaying exact periodic orbits}

\author{Nicolaos D. Caranicolas \and Euaggelos E. Zotos}

\institute{Nicolaos D. Caranicolas:
\at Department of Physics, \\
Section of Astrophysics, Astronomy and Mechanics, \\
Aristotle University of Thessaloniki \\
GR-541 24, Thessaloniki, Greece \\
\and
Euaggelos E. Zotos: \\\email{evzotos@astro.auth.gr}
}

\date{Received: 29 November 2011 / Accepted: 22 February 2012 / Published online: 17 March 2012}

\titlerunning{Investigating the nature of motion in 3D perturbed elliptic oscillators displaying exact periodic orbits}

\authorrunning{Nicolaos D. Caranicolas \& Euaggelos E. Zotos}

\maketitle

\begin{abstract}

We study the nature of motion in a 3D potential composed of perturbed elliptic oscillators. Our technique is to use the results obtained from the 2D potential in order to find the initial conditions generating regular or chaotic orbits in the 3D potential. Both 2D and 3D potentials display exact periodic orbits together with extended  chaotic regions. Numerical experiments suggest, that the degree of chaos increases rapidly, as the energy of the test particle increases. About $97\%$ of the phase plane of the 2D system is covered by chaotic orbits for large energies. The regular or chaotic character of the 2D orbits is checked using the $S(c)$ dynamical spectrum, while for the 3D potential we use the $S(c)$ spectrum, along with the $P(f)$ spectral method. Comparison with other dynamical indicators shows that the $S(c)$ spectrum gives fast and reliable information about the character of motion.

\end{abstract}

\section{Introduction}

It has been almost five decades, since the pioneer work of H\'{e}non and Heiles [1], who studied the regular and chaotic character of motion in a system of two coupled harmonic oscillators. During these years several methods have been presented in order to distinguish order from chaos. Among them, it is worth mentioning, the method based on the Fourier spectra of trajectories (see [2,18]), the Lyapunov Characteristic Exponents (L.C.Es)(see [19]) and the KS entropy (see [3]).

Apart from the above methods, in systems of two degrees of freedom, one can apply the classical method of the Poincar\'{e} phase plane, in order to distinguish regular from chaotic motion. On the contrary, in a 3D system the above method can not be applied. On this basis, it is of particular interest to locate the regions of initial conditions in 3D potentials generating regular or chaotic motion.

The aim of the present article is to investigate the nature of motion in the 3D potential
\begin{eqnarray}
V(x,y,z) &=& \frac{\omega^2}{2} \left(x^2+y^2+z^2\right) \nonumber \\
&+& \epsilon \left[x^2y^2+y^2z^2+x^2z^2-x^2y^2z^2\right], \ \ \
\end{eqnarray}
where $\omega$ is the common frequency of oscillation along $x,y$ and $z$ axis, while $\epsilon$ is the perturbation strength. Potential (1) represents three coupled harmonic oscillators in the 1:1:1 resonance. Potentials of this type are also known as perturbed elliptic oscillators (see [5,6]). The basic reason for the choice of potential (1) is that perturbed elliptic oscillators appear very often in galactic dynamics and atomic-particle physics (see [5] and references therein). A second reason is that it displays exact periodic orbits, interesting sticky orbits together with large chaotic regions. Therefore, it gives a good chance in order to check the reliability of the $S(c)$ spectrum (see [7,10]), which is used as an indicator to track the sticky orbits and to distinguish between regular and chaotic motion.

The outcomes of this research are mainly based on the numerical integration of the equations of motion
\begin{eqnarray}
\ddot{x} &=& -\frac{\partial \ V(x,y,z)}{\partial x}, \ \ \ \nonumber \\
\ddot{y} &=& -\frac{\partial \ V(x,y,z)}{\partial y}, \ \ \ \nonumber \\
\ddot{z} &=& -\frac{\partial \ V(x,y,z)}{\partial z}, \ \ \ \nonumber \\
\end{eqnarray}
where the dot indicates derivatives with respect to the time.

The Hamiltonian to the potential (1) reads
\begin{equation}
H=\frac{1}{2}\left(p_x^2+p_y^2+p_z^2\right)+V(x,y,z)=h \ \ \ ,
\end{equation}
where $p_x,p_y$ and $p_z$ are the momenta per unit mass conjugate to $x,y$ and $z$, while $h$ is the numerical value of the Hamiltonian.

The paper is organized as follows: In Section 2 we present the analysis of the structure of the $\left(x, p_x\right)$ Poincar\'{e} phase plane of the 2D system for a set of values of energy $h$. In the same Section, we give some examples for the reader to understand the usefulness of the $S(c)$ spectrum, since it has the ability to describe islandic and sticky motion. In Section 3 we study the character of orbits in the 3D system. Special interest is given to the evolution of the exact periodic orbits. Particularly, we are interested to find the value of energy, where orbits starting near the exact periodic orbits change their character from regular to chaotic. We close our investigation with Section 4, where a discussion and the conclusions of this research are presented.

\section{Order and chaos in the 2D system}

In the following we shall study the character of motion in the corresponding 2D potential
\begin{equation}
V(x,y)=\frac{\omega^2}{2} \left(x^2+y^2\right) + \epsilon x^2y^2 \ \ \ .
\end{equation}
It is well known, that in this potential the two axes $x$ and $y$ and the straight lines $x=\pm y$ are exact periodic orbits (see [8]). The Hamiltonian to potential (4) writes
\begin{eqnarray}
H_2 &=& \frac{1}{2}\left(p_x^2+p_y^2\right)+V(x,y) \nonumber \\
 &=& \frac{1}{2}\left(p_x^2+p_y^2+ \omega ^2 x^2+ \omega ^2 y^2 \right)+\epsilon x^2y^2 \nonumber \\
 &=& h_2 \ \ \ ,
\end{eqnarray}
where $h_2$ is the numerical value of the Hamiltonian. It is interesting to note, that potential (4) has not a finite energy of escape, so the corresponding zero velocity curves (ZVCs) are always closed. In all numerical calculations, we adopt the values: $\omega =1, \epsilon =1$, while the energy is treated as a parameter.

Figure 1a shows the $\left(x, p_x\right)$, $(y=0, p_y>0)$ phase plane, when $h_2=1$. One can see, that a large part of the phase plane is covered by chaotic orbits. The regular regions are occupied by invariant curves around the points $\left(x_0, p_{x0}\right)=(0, 0)$ and $\left(x_0, p_{x0}\right)=\left(0, \pm \sqrt{h}\right)$. The above points give the positions of two exact periodic orbits, The first is the $y$ axis while the second describes the $x=\pm y$ straight line orbits on the $\left(x, p_x\right)$ phase plane. Furthermore, one observes a large number of smaller islands of invariant curves produced by secondary resonances and some sticky regions as well. Figure 1b is similar to Fig. 1a but when $h_2=2$. Here we see that the majority of the phase plane is covered by a chaotic sea. The regular regions are confined only around the straight line periodic orbits, while the $y$ axis has now become unstable. Some smaller islands of invariant curves are also present. The most interesting feature observed in this case, is the sticky regions around the chain of islands of invariant curves on the $x$ as well as on the $p_x$ axis. We shall come to this point later in this Section. If we increase the value of energy to $h_2=4$, we obtain the phase plane shown in Figure 1c. In this case, the $y$ axis has returned to stability, while the straight line periodic orbits are unstable. Almost all the phase plane is chaotic, except of a small region around the center and some small islands of invariant curves which are products of secondary resonances. A sticky region around the chain of the small islands of invariant curves near the center is also observed. Figure 1d is similar to Fig. 1a but when $h_2=7.174235$. This value of the energy is the value of the energy of escape of the 3D potential (see next Section). Here almost all the phase plane is covered by chaotic orbits. There is a tiny regular region near the center. A careful observation also shows some very small regular islands of invariant curves embedded in the chaotic sea.
\begin{figure*}[!tH]
\resizebox{\hsize}{!}{\rotatebox{0}{\includegraphics*{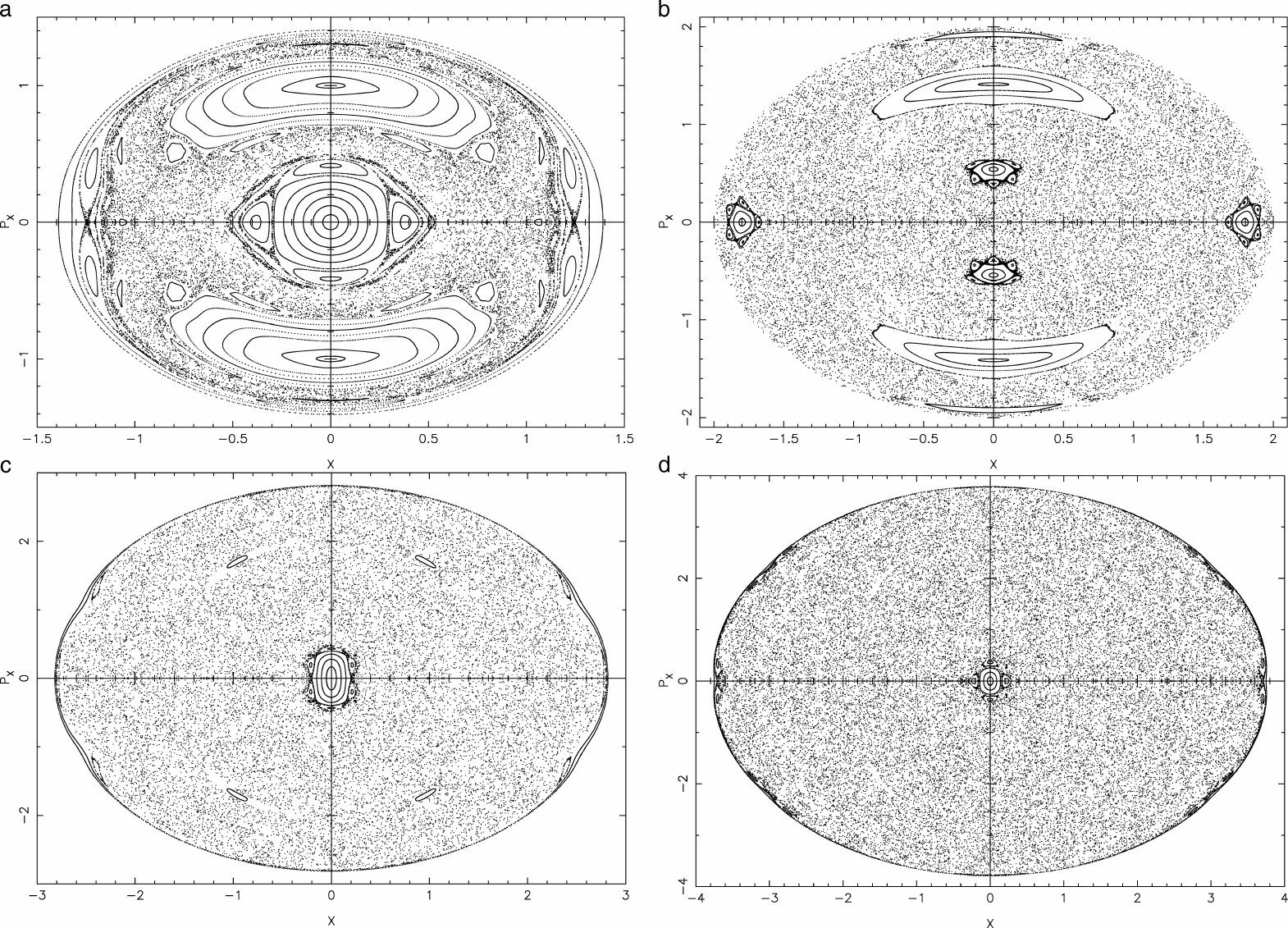}}}
\vskip 0.02cm
\caption{(a-d): The $\left(x, p_x\right)$ phase plane, when $\omega =1, \epsilon =1$, while (a-\textit{upper left}): $h_2=1$, (b-\textit{upper right}): $h_2=2$, (c-\textit{down left}): $h_2=4$ and (d-\textit{down right}): $h_2=7.174235$.}
\end{figure*}

One can see, that the fraction of the phase plane covered by chaotic orbits increases as the energy $h_2$ increases. Figure 2a shows a plot of the percentage of the phase plane $A\%$ covered by chaotic orbits vs $h_2$. We observe that $A\%$ increases rapidly, as $h$ increases. Dots indicate values of $A\%$ found numerically, while the solid line is a fourth degree polynomial fitting curve. We would like to make clear, that $A\%$ is estimated on a completely empirical basis, by measuring the area on the $\left(x, p_x\right)$ phase plane occupied by chaotic orbits (see [11]). In order to have an estimation of the degree of chaos from another point of view, we have plotted the maximum L.C.E vs $h_2$. The results are shown in Figure 2b. Note that the L.C.E increases linearly with $h_2$. Here, we must point out that it is well known that the L.C.E has different values in each chaotic component (see [20]). Since we have regular regions in all cases and a large chaotic area with small sticky regions embedded in each chaotic sea, we calculate the average value of the L.C.E in each case by taking fifty orbits with different initial conditions $\left(x_0,p_{x0}\right)$, in each chaotic domain. In all cases, the obtained values of the L.C.Es were different in the fourth decimal point, in the same chaotic area.
\begin{figure*}[!tH]
\centering
\resizebox{\hsize}{!}{\rotatebox{0}{\includegraphics*{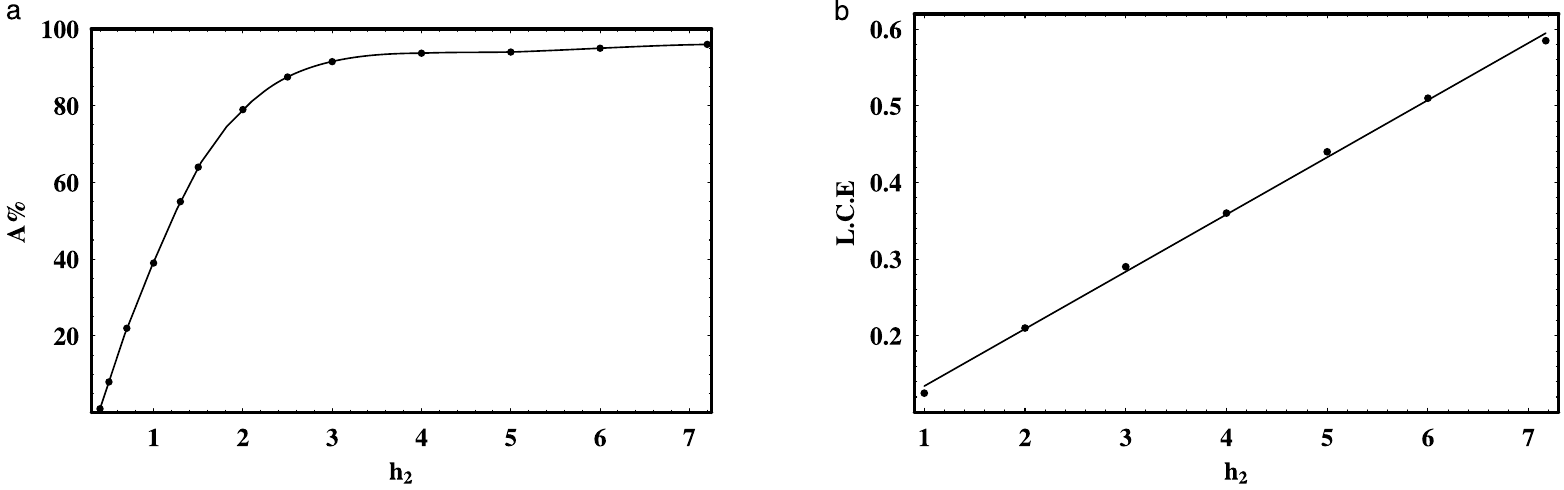}}}
\vskip 0.02cm
\caption{(a-b): (a-\textit{left}) A plot of $A\%$ vs $h_2$ and (b-\textit{right}) a plot of L.C.E vs $h_2$.}
\end{figure*}
\begin{figure*}[!tH]
\centering
\resizebox{0.90\hsize}{!}{\rotatebox{0}{\includegraphics*{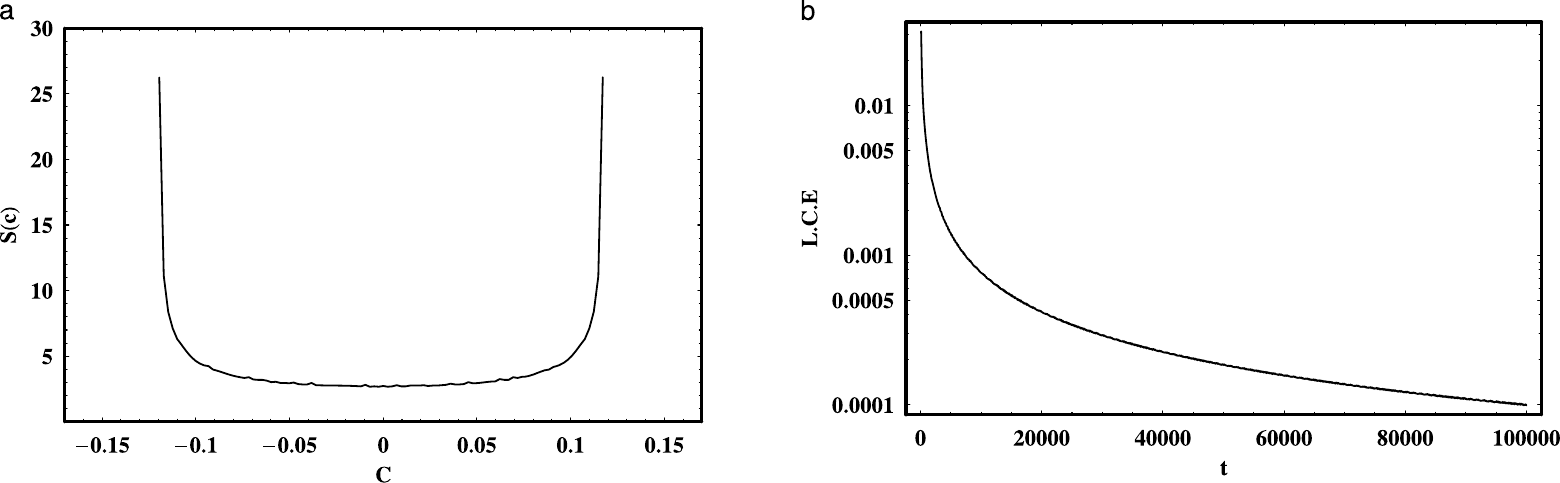}}}
\vskip 0.02cm
\caption{(a-b): (a-\textit{left}) The $S(c)$ spectrum of a regular orbit and (b-\textit{right}) the corresponding L.C.E.}
\end{figure*}
\begin{figure*}[!tH]
\centering
\resizebox{0.90\hsize}{!}{\rotatebox{0}{\includegraphics*{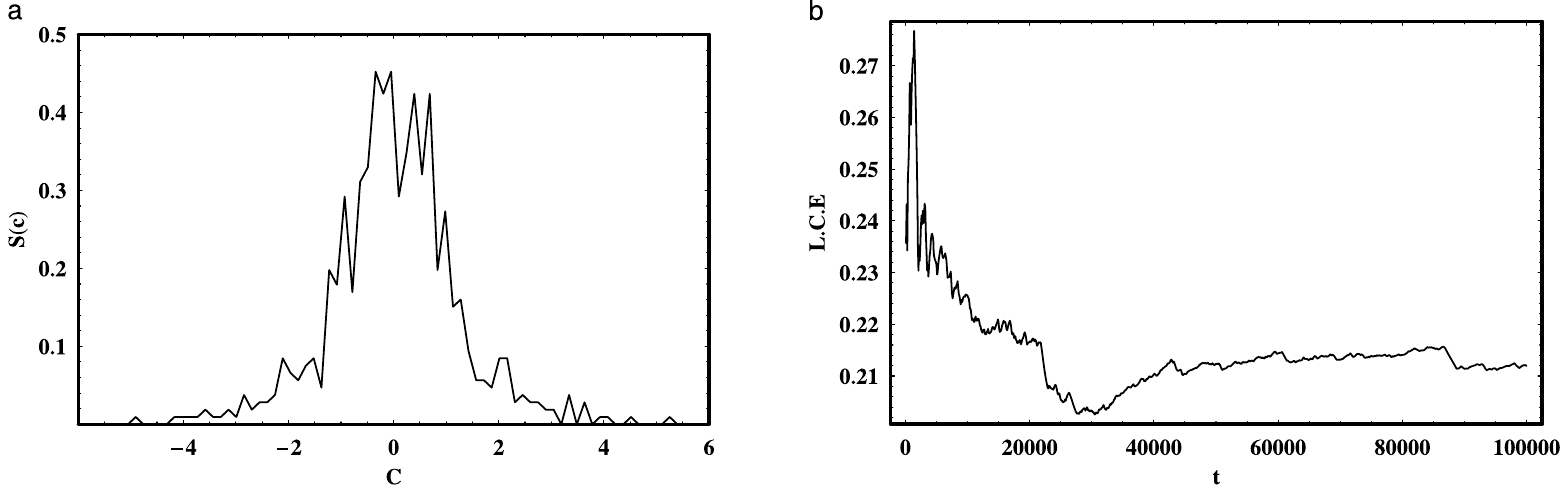}}}
\vskip 0.02cm
\caption{(a-b): (a-\textit{left}) The $S(c)$ spectrum of a chaotic orbit and (b-\textit{right}) the corresponding L.C.E.}
\end{figure*}
\begin{figure*}[!tH]
\centering
\resizebox{0.90\hsize}{!}{\rotatebox{0}{\includegraphics*{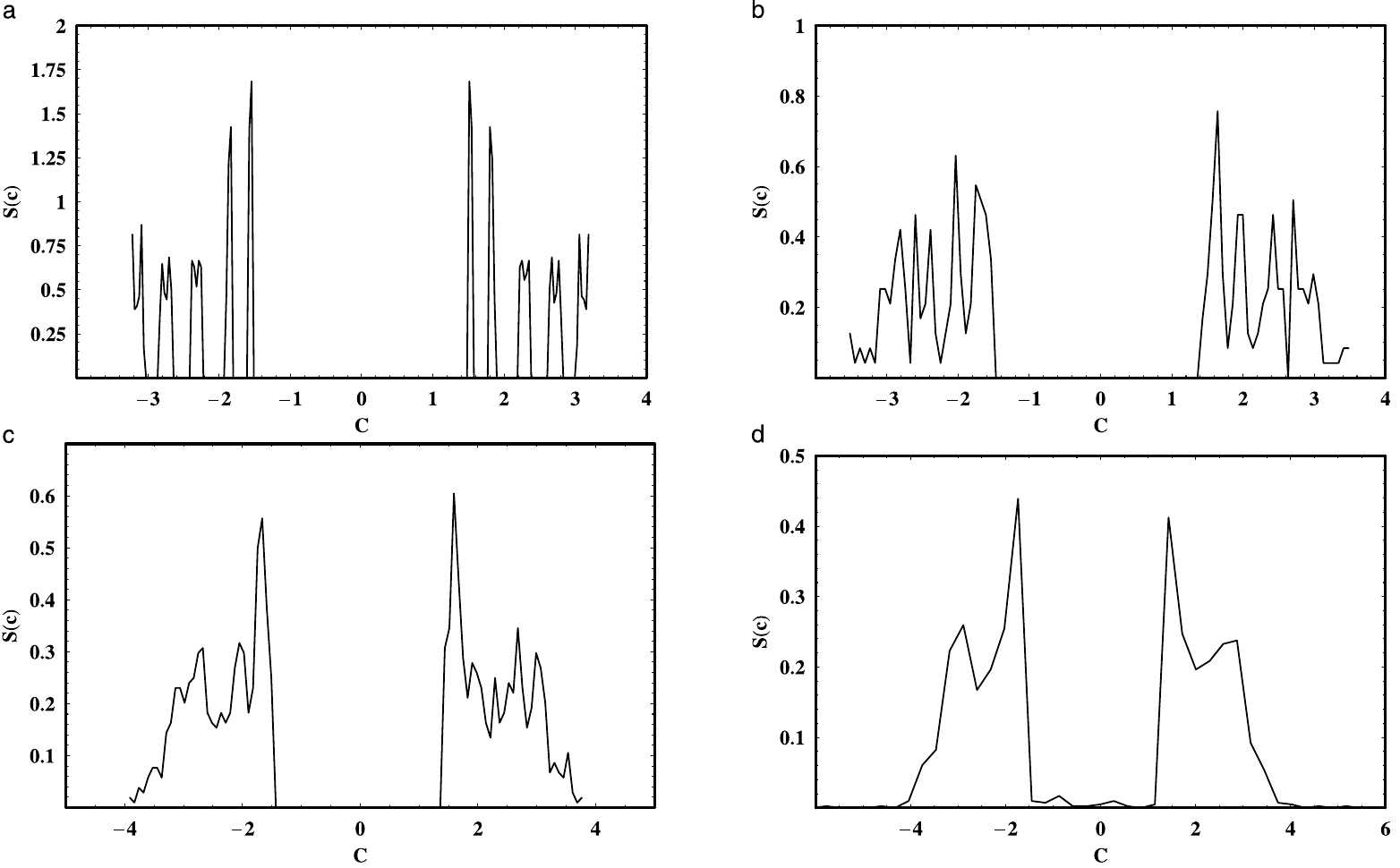}}}
\vskip 0.02cm
\caption{(a-d): (a-\textit{upper left}): The $S(c)$ spectrum of the regular orbit producing the two sets of five islands of invariant curves in the $\left(x, p_x\right)$ phase plane. (b-d): The evolution of the $S(c)$ spectrum of a sticky orbit when (b-\textit{upper right}): $T=1000$, (c-\textit{down left}): $T=4200$ and (d-\textit{down right}): $T=4300$.}
\end{figure*}

Interesting information not only for the chaotic or regular character of orbits but also for islandic and sticky motion can be obtained using the dynamical parameter of the $S(c)$ spectrum. The parameter $c$ is defined as
\begin{equation}
c_i=\frac{x_i-p_{xi}}{p_{yi}} \ \ \ ,
\end{equation}
where $\left(x_i,p_{xi},p_{yi}\right)$ are the successive values of the $\left(x,p_x,p_y\right)$ orbital elements on the Poincar\'{e} $\left(x, p_x\right)$ phase plane. The dynamical spectrum of the parameter $c$ is its distribution function
\begin{equation}
S(c)=\frac{\Delta N(c)}{N \Delta c} \ \ \ ,
\end{equation}
where $\Delta N(c)$ are the number of the parameters $c$ in the interval $\left(c, c+ \Delta c\right)$, after $N$ iterations. This dynamical spectrum has been proved a very reliable tool in several cases (see [7,10] for details). The nature of a 2D orbit can be identified by the shape of the $S(c)$ spectrum. If the structure of the spectrum is a well formed $U$-type shape, then the corresponding orbit is regular. On the other hand, if the shape of the spectrum is complicated, asymmetric, with a lot of of small and large abrupt peaks, then this indicates that the orbit is chaotic. Moreover, the $S(c)$ spectrum can help us identify resonant orbits or orbits of higher multiplicity, as it produces as much $U$-type spectra as the total number of the islands of invariant curves in the $\left(x, p_x\right)$ phase plane. One additional advantage of this spectrum, is that it can be used in order to calculate the sticky period of a 2D sticky orbit and also to help us follow its time evolution towards the chaotic sea.

At this point, we should emphasize that there are some spectra, such as the $S(\alpha)$ spectrum, (see [23]), which have no ability to identify islandic or sticky motion. Furthermore, the $S(\alpha)$ spectrum needs the computation of two neighboring orbits and a large number of iterations (about $5 \times 10^{4}$) in order to provide reliable results regarding the character of a 2D orbit. Using the $S(c)$ spectrum we have managed to overcome these drawbacks. Recently, in [24] we introduced the $S(g)$ spectrum which proved to be a very sensitive and reliable spectrum. This new spectrum definition allow us to study more accurately the islandic motion of resonant orbits of higher multiplicity, as it produces as much spectra as the total number of the islands of invariant curves in the $\left(x, p_x\right)$ phase plane. The main advantage of the $S(g)$ is that it has the ability to detect complicated resonant orbits especially in 2D Hamiltonian systems, no matter how small are the corresponding islands of invariant curves. Furthermore, for the study of orbits in 3D Hamiltonian systems, we constructed and used in [25] the $S(w)$ dynamical spectrum. This spectrum is an advanced form of the $S(c)$ spectrum and it was especially designed for the study of 3D orbits, as it produces as much spectra as the total number of the 3D invariant tori in the $\left(x, p_x, z\right)$ phase subspace.

In what follows we present some examples in order to visualize the usefulness of the $S(c)$ spectrum. Figure 3a-b shows the $S(c)$ spectrum  and the corresponding L.C.E vs time of an orbit with initial conditions: $x_0=0.2, y_0=0, p_{x0}=0$ and $h_2=1$, while in all cases the initial value of $p_{y0}$ is always found from the energy integral. We see a $U$ type spectrum indicating regular motion while the corresponding L.C.E tends asymptotically to zero. Figure 4a-b is similar to Fig. 3a-b for an orbit with initial conditions: $x_0=1.0, y_0=0, p_{x0}=0$ and $h_2=2$. Here we see a spectrum with a lot of asymmetric small and large peaks indicating chaotic motion. The plot of the L.C.E vs time in Fig. 4b verifies the results of the spectrum.
\begin{figure}[!tH]
\centering
\resizebox{\hsize}{!}{\rotatebox{0}{\includegraphics*{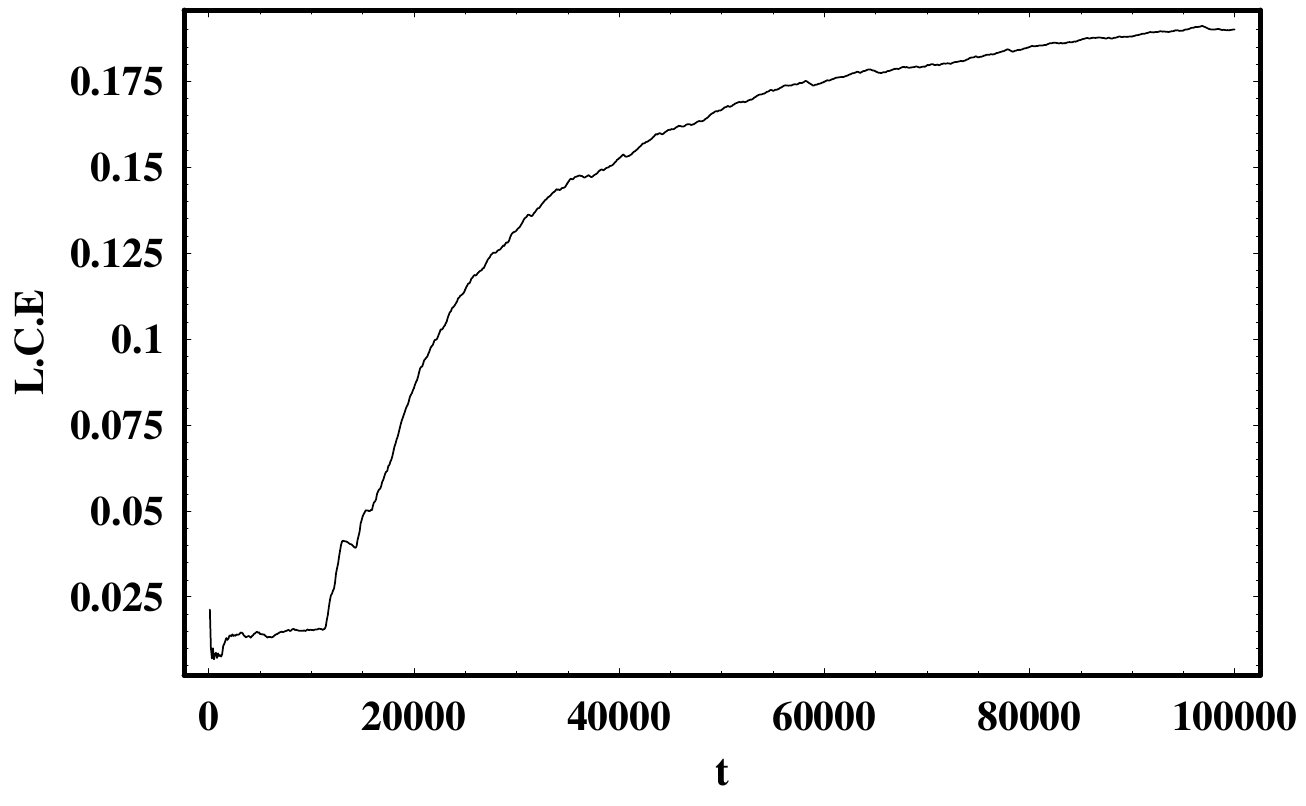}}}
\caption{The L.C.E vs time for the sticky orbit of Fig. 5.}
\end{figure}
\begin{figure}[!tH]
\centering
\resizebox{\hsize}{!}{\rotatebox{0}{\includegraphics*{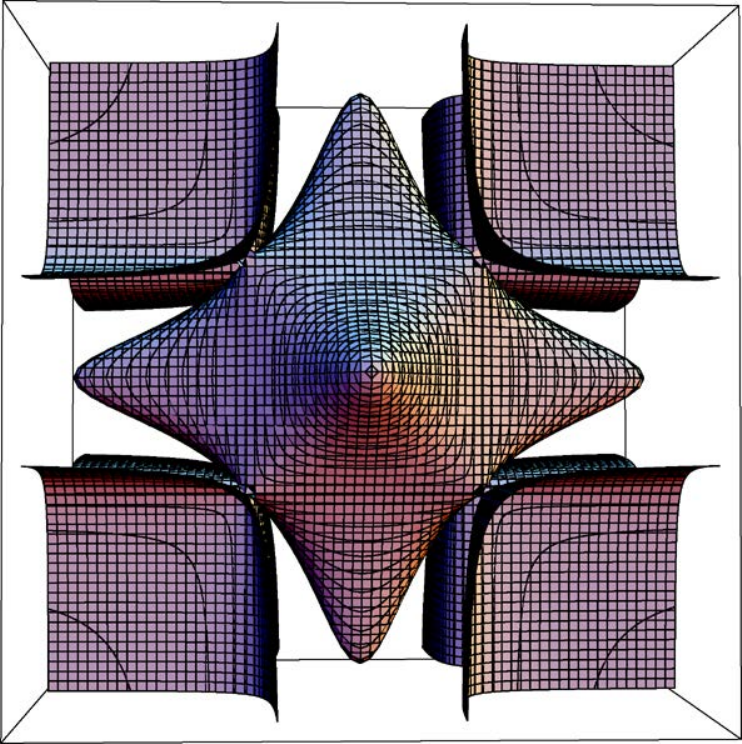}}}
\caption{A plot of the surface $V(x,y,z)=h_{esc}$.}
\end{figure}

The $S(c)$ spectrum is useful in order to detect islandic and sticky motion. The results are shown in Figure 5a-d. Figure 5a shows an orbit with initial conditions: $x_0=-1.682, y_0=0, p_{x0}=0$ and $h_2=2$. This orbit produces the two sets of five islands shown near the $x$-axis in Fig. 1b. Note that the corresponding spectrum produces ten individual regular spectra each one corresponding to an island. As the orbit produces two different sets of five islands on the positive and negative side of the $x$-axis, we observe two sets of five spectra almost symmetric to the origin. Figure  5b shows the spectrum of a nearby orbit. The initial conditions are: $x_0=-1.705, y_0=0, p_{x0}=0$ and $h_2$=2. This orbit produces the two sticky regions around each set of the five islands near the $x$-axis shown in Fig. 1b. As there are two areas of sticky motion we see two separate  sticky spectra with asymmetric large and small peaks. Here the time is $T=1000$ time units. Figure 5c is similar to 5b but when $T=4200$ time units. We see two different asymmetric spectra. This indicates that the test particle is still in the sticky region. Figure 5d shows the spectrum when $T=4300$ time units. In this case, we see that the two spectra have joined to produce a unified chaotic spectrum. This indicates that the test particle has left the sticky region and has gone into the chaotic sea. The L.C.E vs time of the orbit described above is shown in Figure 6. All L.C.Es were computed for a time period of $10^5$ time units.

The above results strongly suggest that the $S(c)$ spectrum is a powerful dynamical parameter providing useful and reliable results for distinguishing between order and chaos. Furthermore, the $S(c)$ spectrum describes very satisfactorily islandic and sticky motion.

\section{The nature of motion in the 3D model}

In this Section we shall investigate the regular or chaotic behavior of the orbits in the 3D potential (1). Working in a similar way as in Caranicolas and Varvoglis [9], we find that this potential has an energy of escape given by
\begin{equation}
h_{esc}=2\epsilon +\frac{3 \omega ^2}{2}+ \sqrt{2\epsilon \left(2\epsilon + \omega ^2 \right)} + \omega ^2
\sqrt{\frac{2\epsilon + \omega ^2}{2\epsilon}}. \ \ \
\end{equation}
For values of $h>h_{esc}$ the zero velocity surface is open and the test particle is free to escape to infinity. In this paper we investigate only bounded motion that is we consider always $h \leq h_{esc}$. For the adopted values of the parameters $\omega =1, \epsilon =1$ we find that $h_{esc}=7.174235$. Figure 7 shows the surface $V(x,y,z)=h_{esc}$. Note that in the potential (1) the straight lines $x=\pm y=\pm z$ are exact periodic orbits. Using the origin $x=y=z=0$ as a starting point we find from Hamiltonian (3) that the starting value of the corresponding velocities are $p_{x0}=p_{z0}=\left(\frac{2h}{3}\right)^{1/2}$, while the value of $p_{y0}$ is obtained from the energy integral.
\begin{figure*}[!tH]
\centering
\resizebox{0.70\hsize}{!}{\rotatebox{0}{\includegraphics*{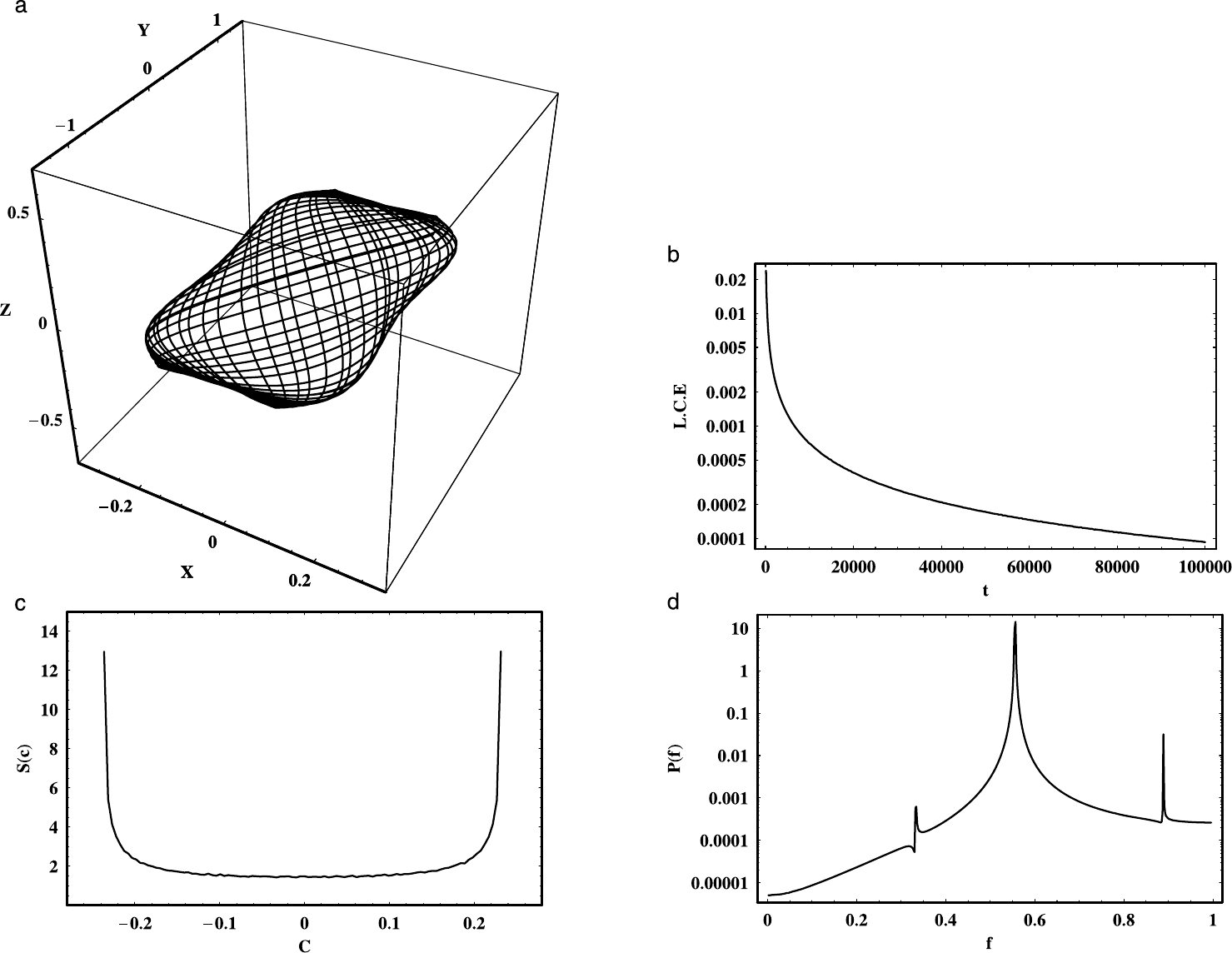}}}
\vskip 0.02cm
\caption{(a-d): (a, \textit{upper left}): A 3D regular orbit, (b, \textit{upper right}): the corresponding L.C.E,
(c, \textit{lower left}): the $S(c)$ spectrum and (d, \textit{lower right}): the $P(f)$ indicator. See text for details.}
\end{figure*}
\begin{figure*}[!tH]
\centering
\resizebox{0.70\hsize}{!}{\rotatebox{0}{\includegraphics*{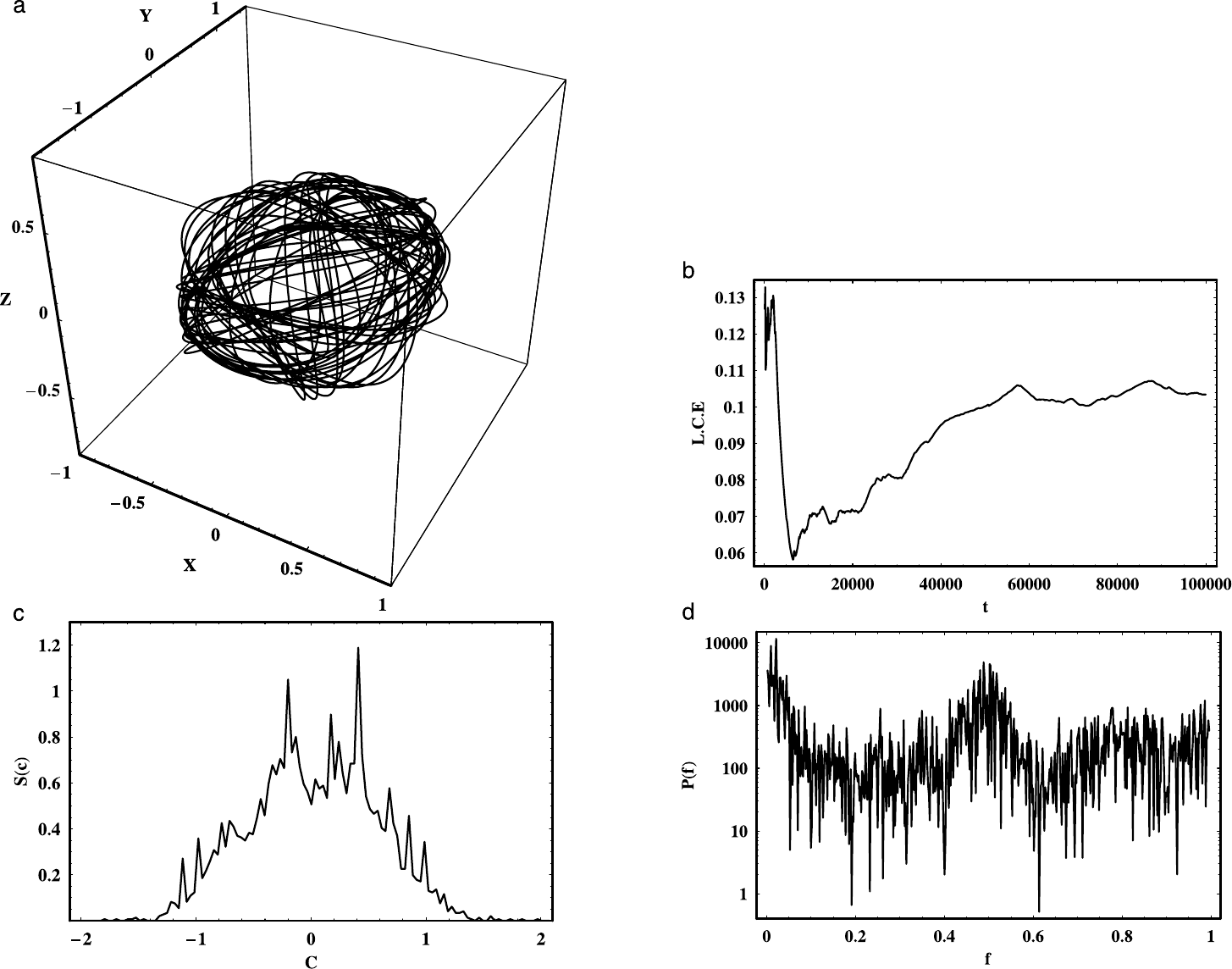}}}
\vskip 0.02cm
\caption{(a-d): (a, \textit{upper left}): A 3D chaotic orbit, (b, \textit{upper right}): the corresponding L.C.E,
(c, \textit{lower left}): the $S(c)$ spectrum and (d, \textit{lower right}): the $P(f)$ indicator. See text for details.}
\end{figure*}
\begin{figure*}[!tH]
\centering
\resizebox{0.70\hsize}{!}{\rotatebox{0}{\includegraphics*{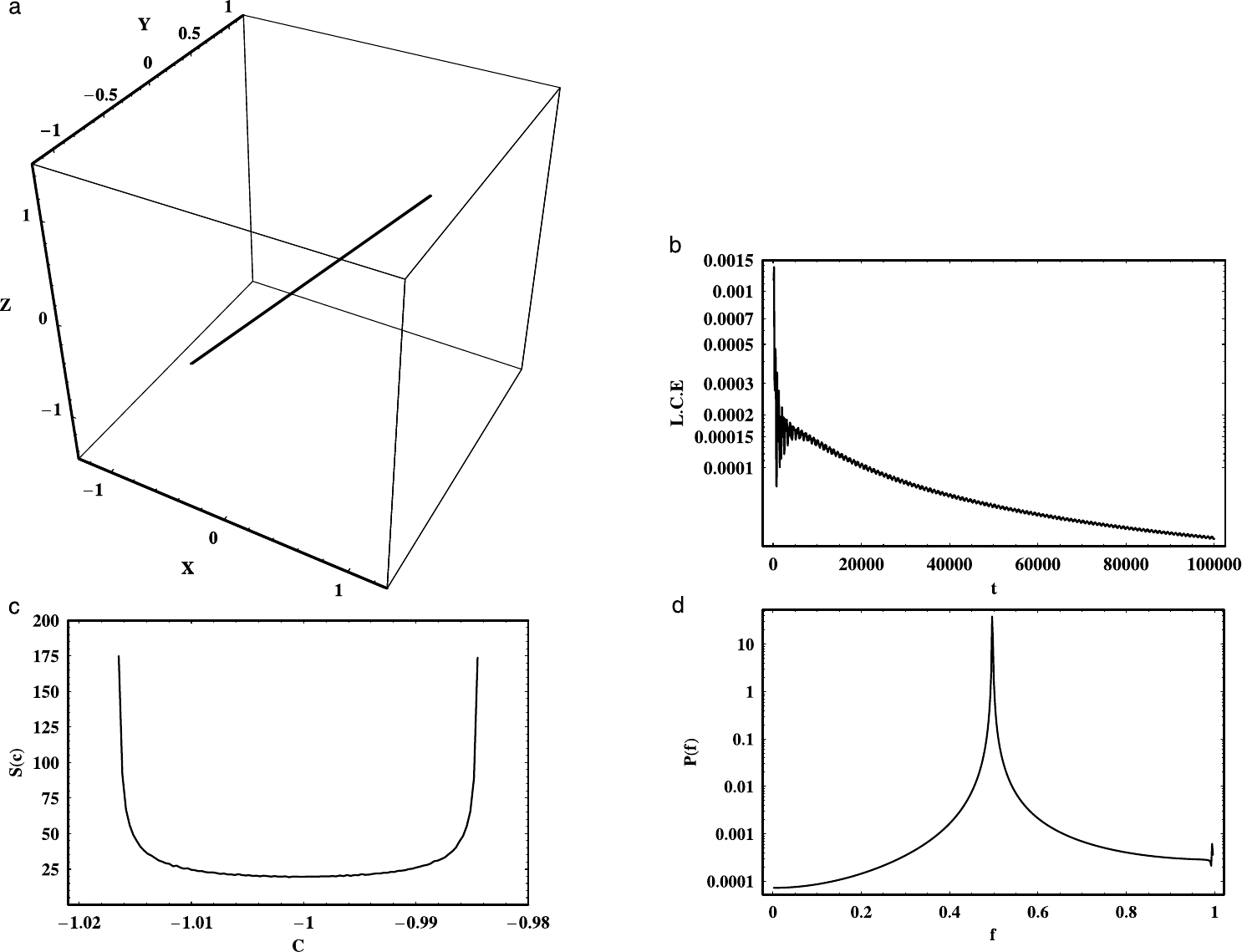}}}
\vskip 0.02cm
\caption{(a-d): Similar to Fig. 8a-d, but for an orbit starting near the exact periodic orbit. The motion is regular for $h=1$.}
\end{figure*}
\begin{figure*}[!tH]
\centering
\resizebox{0.70\hsize}{!}{\rotatebox{0}{\includegraphics*{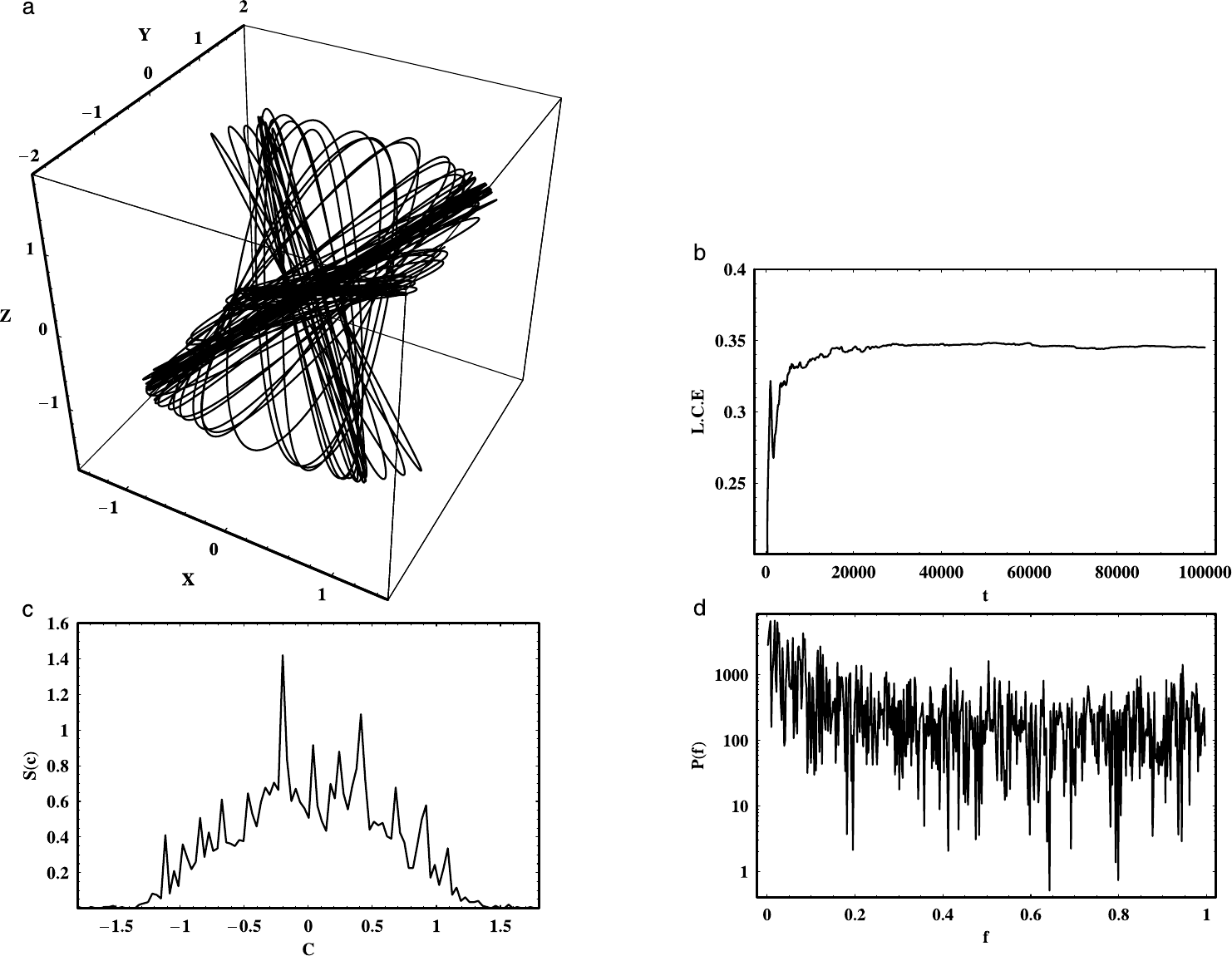}}}
\vskip 0.02cm
\caption{(a-d): Similar to Fig. 9a-d, but for an orbit starting near the exact periodic orbit. The motion is chaotic for $h=5$.}
\end{figure*}

In order to find the regular or chaotic nature of the 3D orbits we choose initial conditions $(x_0,p_{x0},z_0),y_0=p_{z0}=0$, where $(x_0,p_{x0})$ is a point on the phase plane of the 2D system. This point lies inside the limiting curve
\begin{equation}
\frac{1}{2}p_x^2+V(x)=h_2 \ \ \ ,
\end{equation}
which is the curve containing all the invariant curves of the 2D system. We use $h=h_2$ and the value of $p_{y0}$, for all orbits, is obtained from the energy integral (3). The numerical calculations suggest, that orbits with initial conditions $(x_0,p_{x0},z_0),y_0=p_{z0}=0$, such as $(x_0,p_{x0})$ is a point in the chaotic regions of Fig. 1 a-d and for all permissible values of $z_0$, produce chaotic orbits.

On the other hand, it would be also interesting to investigate the character of orbits with initial conditions $(x_0,p_{x0},z_0),y_0=p_{z0}=0$, such as $(x_0,p_{x0})$ is a point in the regular regions of Fig. 1 a-d. We shall use the  $S(c)$ spectrum and in order to verify the effectiveness of the $S(c)$ spectrum in 3D dynamical systems, we shall compare the results with two other indicators, the classical method of the L.C.E and the spectral method $P(f)$, used by Karanis and Vozikis [18]. This method, uses the Fast Fourier Transform (FFT) of a series of time intervals, each one representing the time that elapsed between two successive points on the Poincar\'{e} $\left(x, p_x\right)$ phase plane, for 2D systems, while for 3D systems they take two successive points on the plane $z=0$. Note that the coupling of the third component, $z$, carrying all the information about the 3D orbits, is hidden in the definition of the $S(c)$ spectrum, but in any case it affects the values of $x, p_x$ and $p_y$ entering Eq. (6). Thus, the $S(c)$ spectrum provides implicit results for the 3D orbits. On the other hand, using the new $S(w)$ spectrum the outcomes are explicit, since the coupling of the third component $z$ is located directly in its definition.

Figure 8 a-d shows (a) the 3D orbit, (b) the corresponding L.C.E. (c) the $S(c)$ spectrum and (d) the $P(f)$ indicator, of an orbit with initial conditions: $x_0=0.1,p_{x0}=0,z_0=0.1,y_0=p_{z0}=0$. The value of energy is $h=1$, while in all cases the value of $p_{y0}$ is found from the energy integral. All dynamical indicators coincide that this orbit is regular. Figure 9a-d is similar to Fig. 8a-d but for a chaotic 3D orbit. Initial conditions are: $x_0=0.75,p_{x0}=0,z_0=0.1,y_0=p_{z0}=0$, $h=1$. Here one observes an asymmetric $S(c)$ spectrum with a lot of small and large peaks indicating chaotic motion. The L.C.E in Figure 9b and the $P(f)$ indicator in Figure 9d, coincide with the results of the spectrum.

Finally, we shall study the character of orbits starting near the 3D exact periodic orbits. Figure 10 a-d is similar to Fig. 8 a-d but for an orbit starting near the 3D exact periodic orbit. Initial conditions are: $x_0=y_0=z_0=0,p_{x0}=\left(\frac{2h}{3}\right)^{1/2},p_{z0}=p_{x0}+0.01$, $h=1$. The L.C.E, the $S(c)$ spectrum and the $P(f)$ indicator suggest for regular motion. Figure 11 a-d is similar to Fig. 9 a-d but when $h=5$. Here the dynamical indicators reveal the chaotic nature of this orbit. The numerical calculations indicate that orbits starting near the exact periodic orbits become chaotic when $h \gtrsim 4$. Figure 12 shows a plot of the L.C.E vs $h$ for the chaotic orbits in the 3D system. We see that the L.C.E increases linearly as $h$ increases. Note that the values of the L.C.Es of the 3D system are smaller than the corresponding values of the L.C.Es of the 2D system.

\section{Discussion and conclusions}

Potentials composed of harmonic oscillators have been frequently used during the last years by several researchers [4, 12-16]. In this article we have studied the regular or chaotic character of orbits in a 3D potential made up of perturbed elliptic oscillators. This potential displays exact periodic orbits, that are straight lines going through the origin, together with extended chaotic regions.

Our investigation begins from the 2D potential. In order to keep things simple we tried to study the regular or chaotic character of orbits for fixed values of the two other parameters and different values of the energy. Our numerical calculations have shown that the extend of chaotic regions increases rapidly as the value of the energy increases. On the other hand, we have found that the degree of chaos also increases as the energy of increases. The numerical experiments show that the L.C.E increases linearly with $h_2$. Furthermore, the application of the $S(c)$ spectrum shows that this dynamical parameter is a fast and reliable indicator in order to distinguish order from chaos. Note that in order to obtain the $S(c)$ spectrum we need only one orbit and 3000-10000 time units, while in order to obtain the L.C.E we use two orbits and much longer time periods. Moreover, the $S(c)$ spectrum describes in a very satisfactorily way islandic and sticky motion. On the other hand, the time needed in order to obtain the $P(f)$ indicator, is of order of 5000 time units.

Using the results of the 2D potential we proceeded to study the 3D model. First we tried to find the character of the 3D orbits with $h=h_2$ and initial conditions $(x_0,p_{x0},z_0),y_0=p_{z0}=0$, such as $(x_0,p_{x0})$ is a point in (i) in the chaotic regions or (ii) in the regular regions of Fig. 1 a-d. The numerical results indicate that all orbits of type (i) are chaotic. The orbits of type (ii) display a different behavior. The numerical outcomes suggest that for small values of $z$ $(z \lesssim 0.1)$ the orbits remain regular, while for larger values of $z$ $(z>0.1)$ the motion becomes chaotic.

The regular or chaotic nature of the 3D orbits can be found using the $S(c)$ spectrum. Comparison with the L.C.E and the $P(f)$ spectral method, shows that the $S(c)$ spectrum is a good and fast indicator in order to distinguish regular from chaotic motion, in 3D dynamical systems. In the present research, we have choose to use dynamical parameters, such as the $S(c)$ spectrum and the $P(f)$ indicator, in order to characterize the nature of orbits, because the main advantage of these new methods is that they can provide reliable and conclusive results in short integration time periods, that is $10^{3}-10^{4}$ time units. Here, we must point out that there are also other new dynamical indicators, such as the FLI [17], SALI [21] and GALI [22]. These indicators need at most 500 to 1000 time units of integration time, in order to reveal the true nature of an orbit in Hamiltonian systems. The main disadvantage of the above methods is that we can not use them to identify resonant orbits of higher multiplicity which correspond to multiple islands of invariant curves in the phase plane. Moreover, FLI and GALI are not very sensitive in the case of sticky motion. On the contrary, the SALI method can provide reliable information about sticky orbits, using shorter integration time than the S(c) spectrum.
\begin{figure}[!tH]
\centering
\resizebox{\hsize}{!}{\rotatebox{0}{\includegraphics*{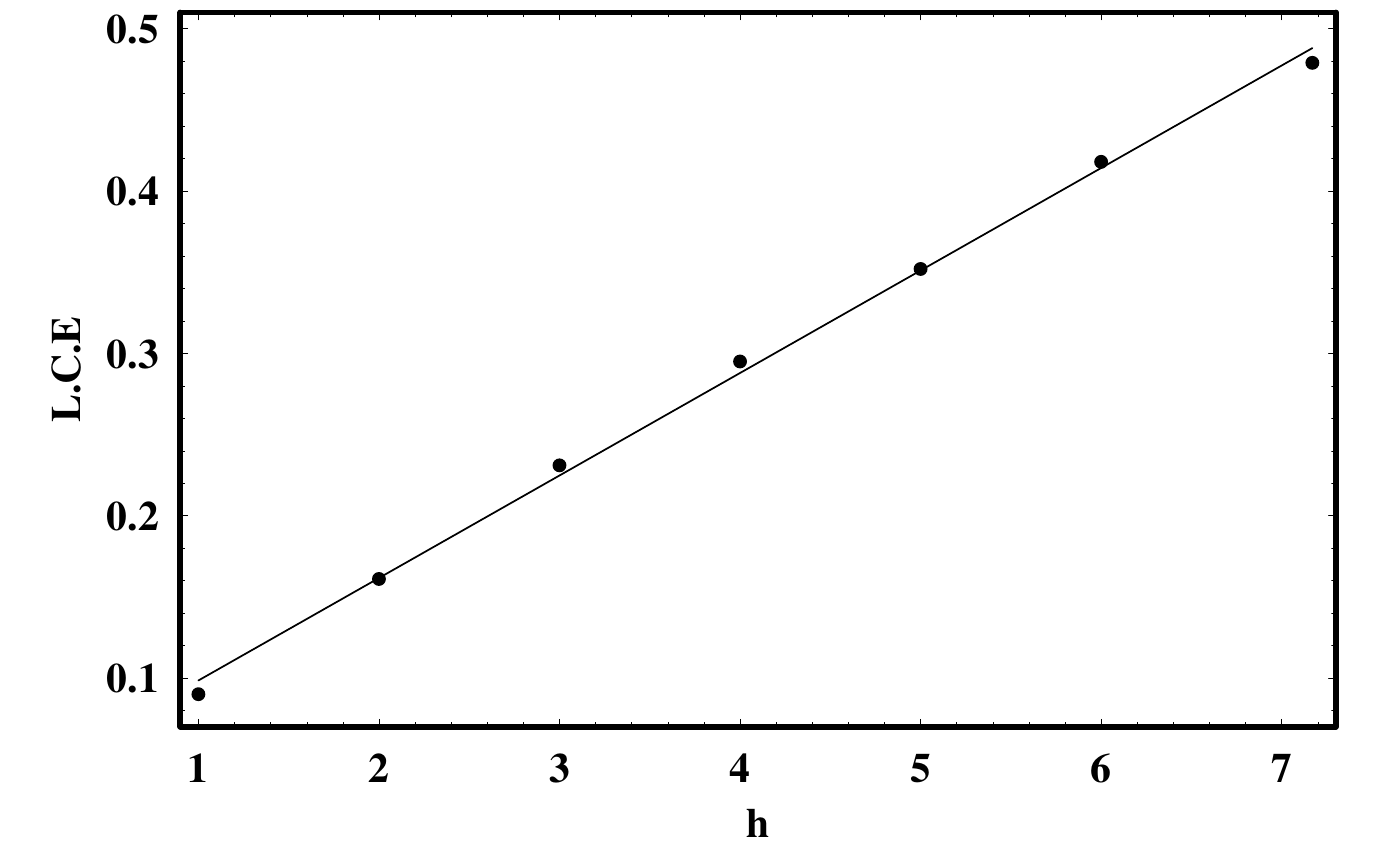}}}
\caption{A plot of L.C.E vs $h$.}
\end{figure}

Note that an interesting characteristic of this investigations is that both, the 2D and 3D potentials display exact periodic orbits and  large chaotic regions. The numerical results, in the 2D and the 3D model, show that, for small values of the energy, orbits starting near the exact periodic orbits remain regular, while for larger values of the energy become chaotic.

\section*{Acknowledgment}

The authors would like to thank the two anonymous referees for their very useful and also constructively suggestions and comments, which improved the quality of the present work.

\end{document}